# Reconfigurable pinwheel artificial-spin-ice and superconductor hybrid device

Yang-Yang Lyu[1,2,+], Xiaoyu Ma[3,+], Jing Xu[4], Yong-Lei Wang[1,*], Zhi-Li Xiao[2,5,*], Sining Dong[1], Boldizsar Janko[3], Huabing Wang[1,6,*], Ralu Divan[4], John E. Pearson[2], Peiheng Wu[1] and Wai-Kwong Kwok[2]

[1]Research Institute of Superconductor Electronics, School of Electronic Science and Engineering, Nanjing University, Nanjing, China

[2]Materials Science Division, Argonne National Laboratory, Argonne, IL, USA

[3]Department of Physics, University of Notre Dame, Notre Dame, IN, USA

[4]Center for Nanoscale Materials, Argonne National Laboratory, Argonne, IL, USA

[5]Department of Physics, Northern Illinois University, DeKalb, IL, USA

[6]Purple Mountain Laboratories, Nanjing, China

+ Authors contribute equally

* Correspondence to: yongleiwang@nju.edu.cn; xiao@anl.gov; hbwang@nju.edu.cn

ABSTRACT: The ability to control the potential landscape in a medium of interacting particles could lead to intriguing collective behavior and innovative functionalities. Here, we utilize spatially reconfigurable magnetic potentials of a pinwheel artificial spin ice structure to tailor the motion of superconducting vortices. The reconstituted chain structures of the magnetic charges in the artificial pinwheel spin ice and the strong interaction between magnetic charges and superconducting vortices allow significant modification of the transport properties of the underlying superconducting thin film, resulting in a reprogrammable resistance state that enables a reversible and switchable vortex Hall effect. Our results highlight an effective and simple method of using artificial spin ice as an in-situ reconfigurable nanoscale energy landscape to design reprogrammable superconducting electronics, which could also be applied to the in-situ control of properties and functionalities in other magnetic particle systems, such as magnetic skyrmions.

Abrikosov vortices are quantized magnetic flux in a type II superconductor, having non-superconducting cores surrounded by swirling supercurrents[1]. As moving vortices dissipate energy, controlling the motion of vortices is essential to tailor superconducting electromagnetic properties for designing innovative electronics[2], such as vortex pumps, diodes, lenses and rectifiers[3]. Vortex dynamics is an active research area, with respect both to fundamental physical studies of vortex matter and to practical electronic and power applications[4]. Vortices can be trapped by nanoscale defects introduced by ion irradiation[5,6], grain boundaries[7,8], nano-strained regions[9], as well as by patterned nanostructures via advanced nanofabrication[10–16]. However, in the aforementioned approach, the spatial vortex pinning potential landscapes produced by defects are fixed, once the samples are synthesized. On the other hand, superconducting vortices can also interact with magnetic nanostructures[17–27]. The controlled magnetization of magnetic nanostructures with an applied magnetic field allows in-situ tuning of the vortex pinning potential, enabling reconfigurable superconducting functionalities[25-27]. Recently, a particular type of magnetic nanostructure, called artificial spin ice (ASI), was used as nanoscale reconfigurable magnetic landscapes to modulate the behavior of vortices in a hybrid ferromagnetic/superconducting structure[28,29]. The highly tunable and wide spectrum of magnetic configurations of artificial spin ice structures allow the tuning of vortex motion and superconducting transport properties in a more flexible way. For example, a reprogrammable vortex diode has been recently realized by patterning a tri-axial artificial spin ice on top of a superconducting film[28], and a reversible vortex ratchet effect was demonstrated by using an artificial kagome spin ice[29]. Here, we present a new artificial-spin-ice and superconductor hetero-structure consisting of a pinwheel (or chiral) ASI array on top of a superconducting film (Figure 1a). The switchable chain-like magnetic charges in the pinwheel ASI lead to a strong modulation of vortex motion, allowing us to achieve switching of energy dissipation and reversal of the vortex Hall effect.

Artificial spin ices contain arrays of interacting single domain nanoscale bar magnets[30]. They were introduced to investigate complex magnetic phenomena, such as geometric frustration. There are various types of ASIs, each containing nanomagnets of different arrangements and exhibiting unique properties[31]. Recently, a pinwheel ASI was created by rotating each of the magnetic nano-bars in a square ASI by 45º, forming an array of orthogonally oriented nanomagnets pointed towards the midpoint of each other[32,33]. The pinwheel ASI forms a four-fold degenerate ferromagnetic order with easily tunable magnetic configurations via an in-plane external magnetic field[33]. The interaction between a nanomagnet in an ASI and a vortex in a superconductor can be viewed as a dumbbell of magnetic charges 'one negative, one positive' attracting and/or repelling a superconducting vortex, respectively[28]. As shown in Figure 1a, one unique feature of pinwheel ASI is the formation of alternating rows of positive and negative magnetic charge chains, creating alternating negative (low) potential valleys and positive (high) potential ridges for superconducting vortices. The orientation of these charge chains can be conveniently rotated 90 degrees by tuning the ASI's magnetization with an in-plane magnetic field (Figures 2b and 2e). This leads to a reconfigurable magnetic potential landscape, enabling the control of vortex motion to realize a reversible superconducting switch and a programmable vortex Hall effect. Our results could stimulate future investigations to explore abilities and advantages of reprogrammable hybrid devices based on reconfigurable magnetic potentials.

Our experiments were carried out on a hetero-structure containing a pinwheel ASI array fabricated on top of a 100 nm thick superconducting $Mo_{0.79}Ge_{0.21}$ (MoGe) film (Figure 1a). An image of the sample is shown in Figure 1b. The superconducting MoGe film was patterned into a micro-bridge containing two separate sections (for demonstrating different effects) using photolithography. Subsequently, two pinwheel ASI arrays, oriented 45 degrees from each other were separately fabricated on top of the MoGe sections via e-beam lithography, followed with e-beam evaporation of permalloy. The permalloy islands have dimensions of 350 nm (l) × 80 nm (w) × 25 nm (d). In each section there are 18,000 nanomagnets at the measured area. Lastly, 100 nm thick gold electrodes were deposited for measuring both regular (longitudinal) and Hall (transverse) voltages. Transport measurements were conducted in a three-axis superconducting vector magnet, which provides magnetic fields in any desired 3D orientation. The applied current is directed along the length of the micro-bridge. An oriented in-plane magnetic field (Figure 1a) is used to tailor the magnetic charge configurations of the ASI and is turned off when the transport measurements are being conducted. During the transport measurements an out-of-plane magnetic field (perpendicular to the sample plane) is applied to tune the density of superconducting vortices in the MoGe film (Figure 1a). The superconducting transition temperature $T_c$ of our MoGe film is 6.9 K. The experiments were carried out at a temperature of 6 K. Additional sample fabrication and experimental details can be found in Ref. [28].

The pinwheel ASI array of section A of the sample is shown in Figure 2a along with the applied current and vortex motion directions. The magnetic charge order can be in-situ switched between vertical charge chains and horizontal charge chains, as shown by the magnetic force microscopy images in Figures 2b and 2e, respectively. The current induced Lorentz force on the vortices is perpendicular to the applied current (Figure 2a). In Figures. 2c and 2f, we respectively show the magnetic field and current dependences of the measured voltages under the vertical and horizontal magnetic chains. The magnetic field is normalized to the first matching field, $B_0$=165.6 Oe, where the density of vortices is equivalent to the density of nanomagnets (detailed calculation of $B_0$ can be found in the supplemental document). That is, at $B=B_0$ the number of superconducting vortices is the same as that of the nanomagnets. The black dashed lines in Figures 2c and 2f highlight the critical current ($I_c$) curves defined by voltage criteria of ±1 µV. We can see a clear difference of current-voltage (*I-V*) characteristics obtained when the magnetic charge chains are vertical and horizontal, respectively, indicating distinct vortex motion behaviors.

For a direct comparison, we extract the critical current curves from Figures 2c and 2f, and plot them in the same graph (Figure 2d). In both cases, the $I_c(B)$ curves show clear peaks/kinks at the magnetic fields of $B_0$ and $0.5B_0$. These are typical features of vortex matching effects induced by dynamical matching of an ordered vortex lattice to the regularly distributed pinning potentials[13-29]. To unveil the microscopic distribution and dynamics of the vortices, we conduct molecular dynamics (MD) simulations (details on the method are described in Ref. [28]). The simulations (Supplementary Figure S1) show ordered vortices at the magnetic fields of $B_0$ and $0.5B_0$. Figure 2d shows that at $B<B_0$ the critical currents are enhanced when the orientation of the magnetic charge chains is switched from vertical to horizontal. This indicates that vortex motion is hindered under the potentials created by the horizontal charge chains. As mentioned earlier, the chain-like magnetic charges produce alternating low potential valleys and high potential ridges. Our MD simulations (Video 1) clearly show that vortices move along low potential valleys induced by the vertical magnetic charge chains, while they are hindered or even arrested when the magnetic charge chains are horizontal. This is because along the low potential valleys (quasi-1D) of the vertical charge ordered structure, the pinning potential is much flatter than that across the alternating low potential valleys and high potential ridges of the horizontal charge ordered structure. Therefore, the vertical magnetic charge chains produce easy vortex flow channels and the horizontal charge chains create large potential barriers that prevent vortex motion. This leads to very large critical currents when the charge chains are horizontal, as compared to the ones induced by the vertical magnetic charge chains.

The switchable and rotatable magnetic charge chains of the pinwheel ASI provide us with a convenient way to in-situ switch on/off the motion of vortices, enabling a superconducting switch for vortex flow. We demonstrate the superconducting vortex-flow switch by alternatingly toggling the ASI's magnetization states between vertical and horizontal charge chains (Figure 2g). To realize such a switch, we bias the current and the magnetic field (out-of-plane) to the values marked by the blue point in Figure 2d. At this point, the value of the measurement current lies in between the high and low critical currents of the horizontal and vertical charge chains, respectively. In this case, we can switch the sample between dissipationless state and vortex flow (resistive) state by toggling the ASI's magnetic charge order by simply rotating the in-plane magnetic field. The top portion of Figure 2g shows the switching on and off of the dissipationless states. From Figure 2d, we also see that the two critical current curves cross at $B=B_0$, leading to a reversal of the critical currents under the vertical and horizontal charge chains at $B>B_0$. At $B=B_0$, the number of vortices equals that of the nanomagnets, so all the vortices are trapped by the attractive magnetic charges. Additional vortices generated at $B>B_0$ are located at interstitial sites. In this case, the vortex-vortex interaction plays an important role. As shown in Video 2, under vertical magnetic charge chains, all the interstitial vortices are immobilized and pinned by neighboring trapped vortices and the repulsive magnetic charges. In contrast, under horizontal charge chains, lines of interstitial vortices first move together horizontally to seek gaps between the peaks (or repulsive charges) in the high potential ridges, and then pass the high potential ridges by squeezing along the gaps, resulting in a much lower critical current. This results in a reversal of the critical current behavior for the two ASI's configurations at low and high magnetic fields. By biasing a current between the two critical current curves at $B>B_0$ (the green point in Figure 2d), we could reverse the polarity of the vortex switch as compared to $B<B_0$ (see the bottom graph of Figure 2g). That is, we can achieve a reconfigurable superconducting vortex-flow switch in which the sequence of the low- and high-resistance states can be reversed by setting the out-of-plane magnetic field from $B<B_0$ to $B>B_0$.

In addition to switching on/off of the vortex motion, the alternating row-by-row attractive potential valleys and repulsive potential ridges provide a perfect system for guided vortex motion. Since vortices favor to move along the attractive potential valleys we can tailor the path of moving vortices to deviate from their nominal direction of the Lorentz driven force. For example, we could design a vortex Hall device in which vortices could have a longitudinal component of the motion under a transverse driving force. To demonstrate this, we fabricate a pinwheel ASI array rotated by 45º in section B of the sample (Figure 1b and Figure 3a). Such a pinwheel ASI array produces diagonally oriented magnetic charge chains whose orientation can be in-situ switched between -45º (Figure 3c) and +45º (Figure 3d). Since the motion of vortices induces a voltage perpendicular to their moving direction, a longitudinal vortex motion results in a transverse (Hall) voltage. This allows us to detect the vortex Hall effect using the same conventional six-probe electrical contact configuration as that used for investigating the electric charge Hall effect (Figure 1b).

We measured the magnetic field (out-of-plane) and current dependences of the Hall voltages at -45º and +45º orientations of ASI's charge chains. The results are shown in Figures 3f and 3g, respectively. Clear Hall voltage signals are observed, indicating the realization of the vortex Hall effect. The Hall signal (Figures 3f and 3g) shows interesting behavior with positive and negative Hall voltage values over the magnetic field and current maps. To further look into the origins of these Hall signals, we carried out MD simulations. Video 3 shows that the vortices move diagonally along the attractive charge order chains (potential valleys) for $B \lesssim B_0$, where the density of vortices is smaller than that of the attractive charges or nanomagnets. Figures 4a and 4b are screenshots of Video 3 (respectively corresponding to the case for -45º and +45º magnetic chains), which show trajectories of the vortex motion. We explain the diagonal vortex motion by drawing the forces on them, as shown in Figures 4a and 4b. The Lorentz force $F_1$, generated by horizontal electrical currents, is in the vertical direction, and the magnetic charge chains produce force $F_2$ with the direction perpendicular to the chains, preventing vortices from moving across the chains. The resultant force, $F_1+F_2$, drives the vortices in a diagonal direction, which results in a horizontal vortex velocity, leading to a measurable Hall voltage signal. In Figure 4 and Video 3, we can see that the horizontal components of the vortex motion are in the opposite directions for the case of -45º and +45º magnetic charge chains, resulting in inverted polarities of the Hall signals (Figures 3f and 3g). Since the corresponding transverse components of vortex velocities are equivalent under both ±45º magnetic charge chain configurations, the longitudinal voltages are identical for the two cases (see Supplemental Figure S2). In the top graph of Figure 3b, we also demonstrate the switching of Hall signal polarity by rotating the in-plane fields to change the orientation of the magnetic charge chains.

Besides reversing Hall signals by tuning ASI's states, the Hall signals can also be reversed by changing the out-of-plane magnetic field. In both Figures 3f and 3g, the sign of the Hall voltage inverts at high magnetic fields as compared to low magnetic fields. This is again attributed to the strong vortex-vortex interactions. Our simulations for high fields (Figures 4c, 4d and Video 4) clearly demonstrate that the horizontal component of vortex motion is in the opposite direction as compared to that at low magnetic fields ($B<B_0$) (Figures 4a, 4b and Video 3). The bottom and top graph of Figure 3b clearly demonstrates a reversal of the polarity of the Hall switching voltage at different magnetic fields. The reversal of the Hall signal with increasing magnetic fields cannot occur for electronic charges, such as electrons or holes and is a unique property of superconducting vortices.

However, similar to the Hall effect arising from electronic charges, the polarity of the vortex Hall voltage signal is also reversed when the current direction is inverted. This is because an inverted current changes the Lorentz force direction, pushing vortices to move in the opposite direction. However, the Hall voltage signal shows a mirror symmetry with respect to the zero magnetic field line, in contrast to the Hall effect from electronic charges, which shows a reversed Hall voltage when the magnetic field is reversed. For superconducting vortices, reversing the magnetic field would flip their polarity and the Lorenz force acting on them, causing the vortices to move in the opposite direction along the chains. Thus, reversal of the magnetic field orientation will not change the polarity of the Hall voltage signal, giving rise to the mirror symmetry with respect to the zero magnetic field line in Figures 3f and 3g. Furthermore, the pinwheel ASI can be converted into a random magnetic configuration (Figure 3e), thereby switching off the vortex Hall effect (Figure 3h). This demonstrates another advantage of our ASI and superconductor hybrid device, where the Hall effect can not only be reversed, but also switched on and off by tuning the ASI's magnetization state.

The pinwheel (or chiral) artificial spin ice with in-situ tunable array of chain-like magnetic charge order exhibits a high degree of electronic tunability, leading to multiple programmable superconducting properties and functionalities, such as switchable dissipationless/resistive states and reversible Hall effects. Furthermore, not only can the vortex Hall effect signal be reversed by flipping the polarity of the applied current, but it also can be inverted by changing the amplitude of the magnetic field. The vortex Hall effect can even be switched on/off by varying the ordering of the magnetic charges. In addition, dynamical vortex behavior could be adjusted by tuning the angle between the direction of the driving current and the orientation of magnetic charge chains, which can be realized by rotating the orientation of the pinwheel ASI on the superconducting film. The performance of the hybrid device may further be improved by fine-tuning the parameters of the pinwheel ASI, such as the dimensions of the nanomagnets and the ASI lattice spacing, which would alter the interaction strength between the magnetic charges and superconducting vortices. This could lead to even more interesting phenomena and physics, such as reconfigurable dynamic vortex Mott transitions[34]. Lastly, the method demonstrated here for tuning vortex motion using a pinwheel ASI could be applied to other topological charges such as magnetic skyrmions[35], providing a universal and convenient approach to engineer new reprogrammable functionalities in magnetic particle systems. The reconfigurable magnetic potential landscape could also provide a perfect platform for tailoring spin-wave channels[36].

**TOC Graphic**

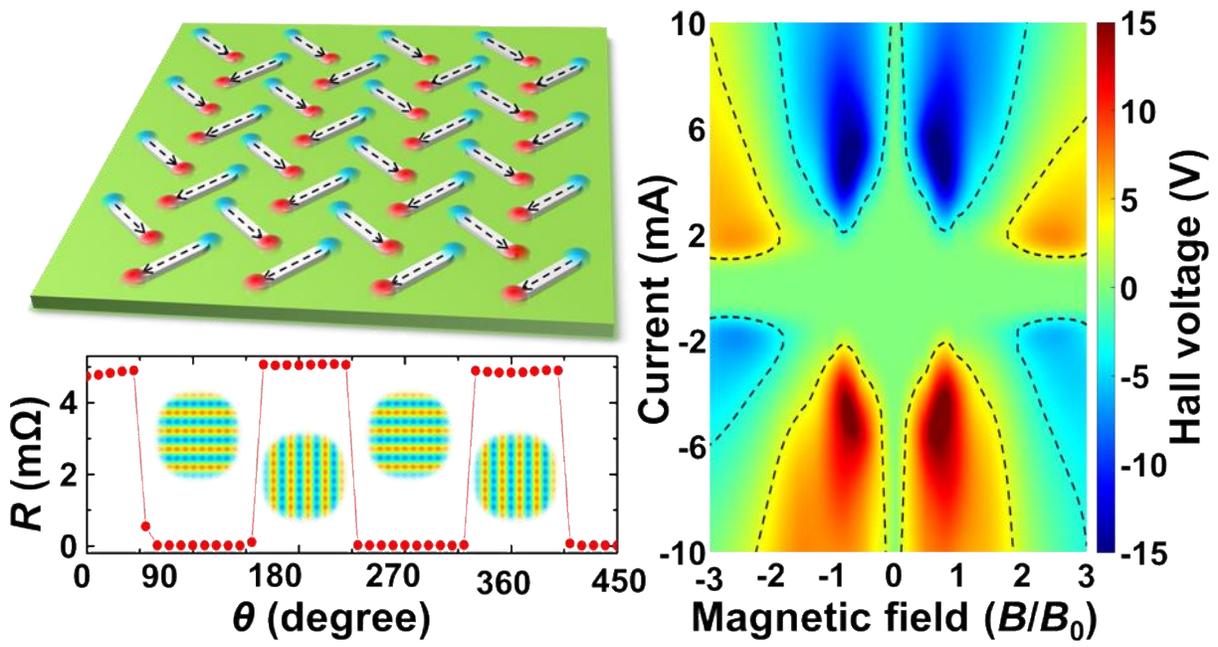

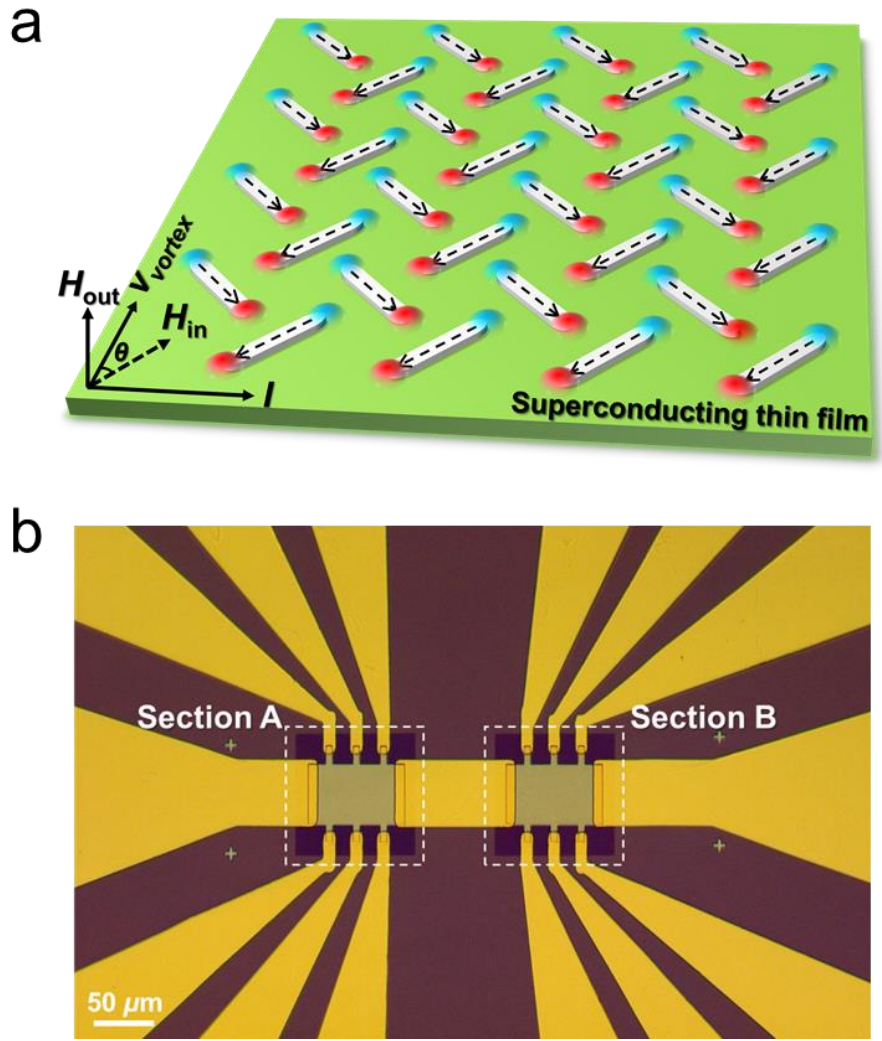

**Figure 1.** Hybrid device of pinwheel artificial spin ice and superconductor. (a) Schematic diagram of the hybrid system consisting of a pinwheel ASI structure on top of a superconducting film. The arrows in the left-lower corner indicate the directions of the applied current, vortex motion, external out-of-plane and in-plane magnetic fields. The dashed arrows on the magnetic nano-bars represent the magnetization directions, and the red and blue spots indicate positive and negative magnetic charges, respectively. (b) Optical image of the hetero-structure sample on a Si/SiO$_2$ substrate. The yellow area is covered by a 100 nm thick layer of gold. The grey rectangles and dark squares show the MoGe micro-bridges and areas covered by the ASI arrays, respectively.

Additional data on Sections A and B including images of the nanomagnets and the magnetic charges as well as the results of transport measurements are presented in Figures 2 and 3, respectively.

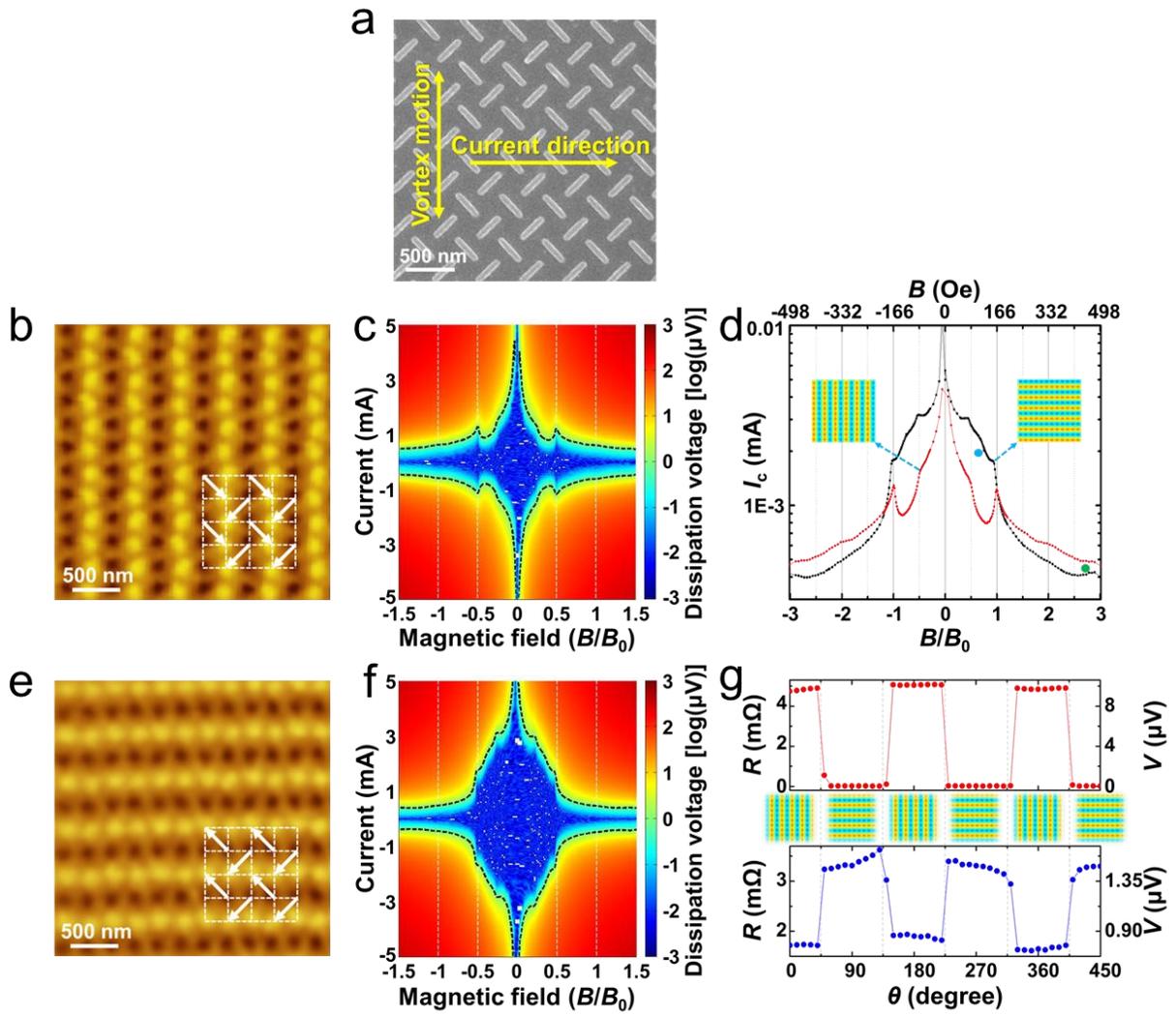

**Figure 2.** Reversible superconducting switch. (a) Scanning electron microscopy image of the pinwheel artificial spin ice in Section A in Figure 1b. (b) and (e) respectively show the magnetic force microscopy images of the vertical and horizontal chains of magnetic charges from the pinwheel ASI shown in (a). The bright and dark spots are positive and negative magnetic charges, respectively. The white arrows represent magnetization directions of the nanomagnets. (c) and (f) show the magnetic field and current dependences of the longitudinal voltages corresponding to the magnetic charge configurations shown in (b) and (e), respectively. The color scale is for the values of the log|$V$|. The black dashed lines indicate the critical currents defined by a voltage criterion of

±1 µV. The magnetic field is in unit of $B_0$=165.6 Oe, corresponding to one vortex per nanomagnet. (d) shows the magnetic field dependence of the critical current $I_c(B)$ curves extracted from (c) and (f). (g) demonstrates the reversible switching of the resistance states. Results shown in the top and bottom graphs in (g) were obtained at $I$=2 mA, $B$=120 Oe and $I$=0.45 mA, $B$=450 Oe, as marked in (d) by the blue and green points, respectively. The angle $\theta$ corresponds to the orientation of the in-plane magnetic field used to polarize the ASI array (see Figure 1). The corresponding magnetic charge configurations are shown in the middle panel.

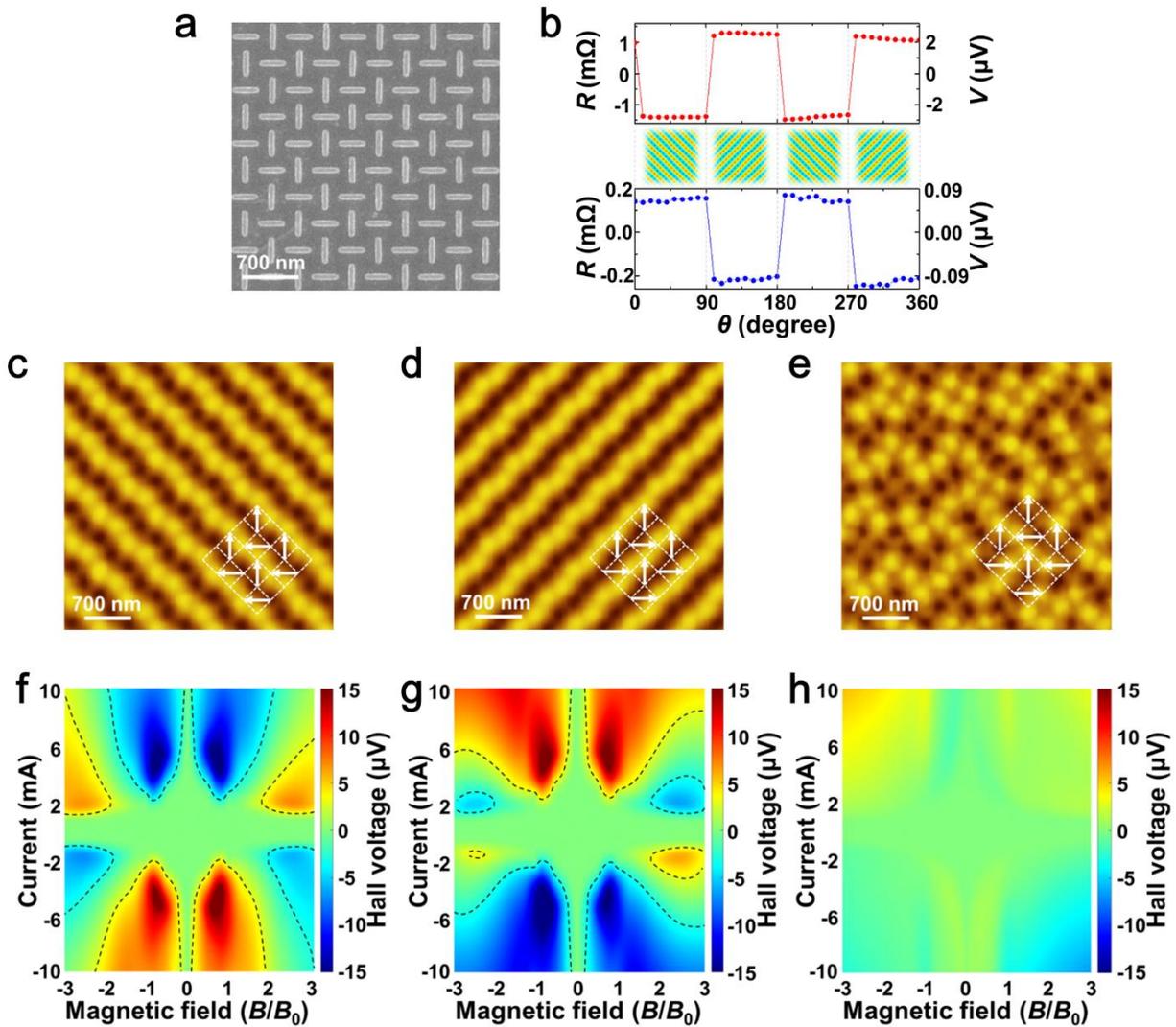

**Figure 3.** Reprogrammable vortex Hall effect. (a) Scanning electron microscopy image of the pinwheel artificial spin ice in Section B in Figure 1b. (b) Reversible switching of the vortex Hall voltages at low magnetic field (120 Oe) (top graph) and at high magnetic field (400 Oe) (bottom graph). (c)-(e) Magnetic force microscopy images of the -45º, +45º magnetic charge chains and a random configuration of magnetic charges obtained from (a). (f)-(h) Magnetic field and current dependences of the vortex Hall voltage respectively, obtained under the magnetic charge configurations shown in (c-e).

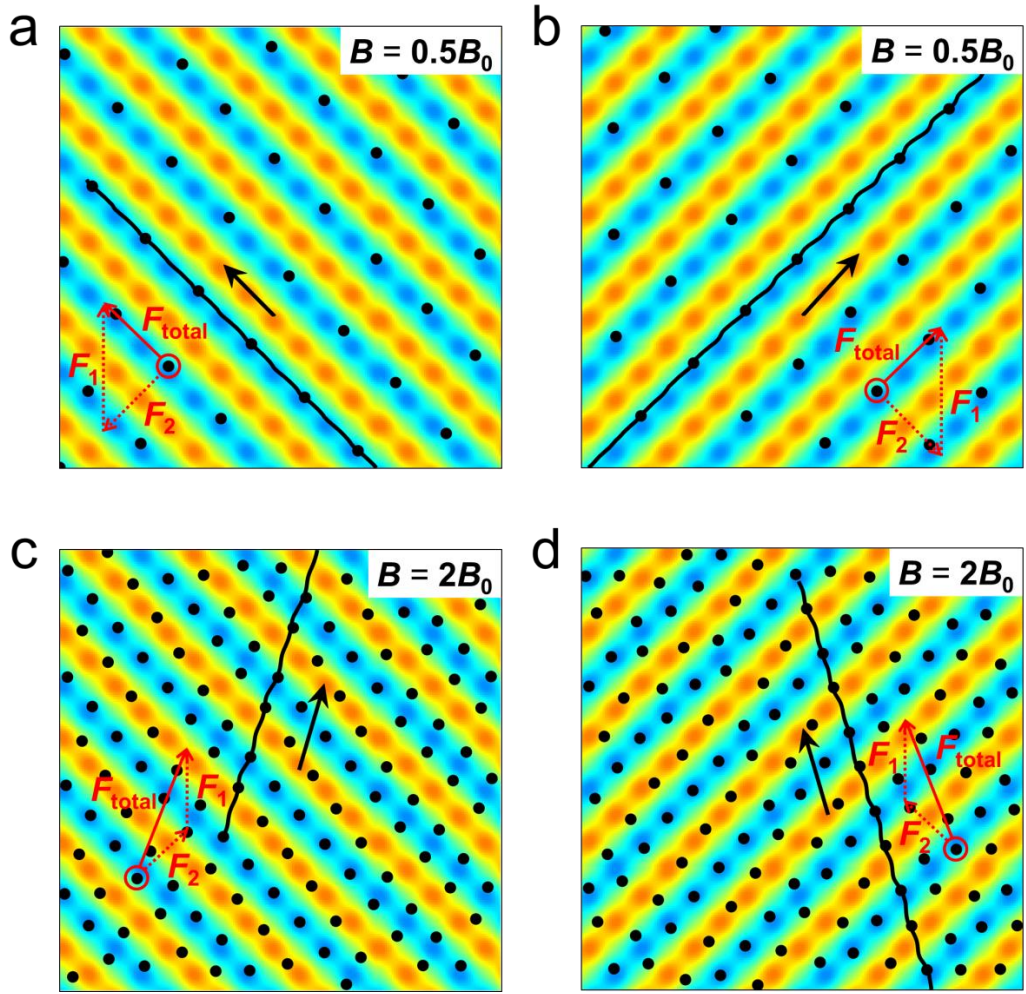

**Figure 4.** Molecular dynamics simulations of the vortex Hall effect. (a)-(d) Screen shots of simulated vortex motions (from Videos 3 and 4) under $B = 0.5B_0$ (a and b) and $B = 2B_0$ (c and d) for -45° and 45° oriented magnetic charge chains. Black dots represent vortices. Blue and red represent attractive and repulsive potentials, respectively. In each figure, the black line is the trajectory for one vortex and the black arrow shows the direction of motion. $F_1$ is the Lorentz driving force induced by the applied current. $F_2$ is the force from magnetic charges. $F_{total}$ is the total force.